\documentclass[12pt]{article}
\usepackage{axodraw}
\usepackage{epsfig}   
\usepackage{amssymb}
\usepackage{latexsym}
\usepackage{amsmath}

\usepackage{graphics,subfigure} 

\usepackage{color}   

\newcommand{\bea}{\begin{eqnarray}}
\newcommand{\eea}{\end{eqnarray}}
\newcommand{\beq} {\begin{equation}}
\newcommand{\eeq} {\end{equation}}

\begin{document}
\pagestyle{empty}
\begin{flushright}
\today
\end{flushright}
\begin{center}
{\large\sc {\bf   Nearly Static Magnetized Kerr Black-hole in Non-linear Electrodynamics }}
\end{center}
\vspace{1.0truecm}
\begin{center}
{\large  K.~G.~Managave,~H.~A.~Redekar,~R.~B.~Kumbhar,~S.~P.~Das \footnote{Email: spd.phy@unishivaji.ac.in} and K.~Y.~Rajpure }\\
\vspace*{5mm}
{}{\it                                                                      
Department of Physics, Shivaji University, \\[0.15cm]
Kolhapur-416004, Maharashtra, India.} \\[0.07cm]
\end{center}

\begin{abstract}
We have analyzed the thermodynamics of slowly rotating magnetized Kerr black-hole, with typical spin parameter $a\le 0.1$ (nearly static) in the background of non-linear electrodynamics. In particular we have studied the Bekenstein-Hawking entropy, Hawking temperature, angular momentum,  specific heats and identified regions of parameters for possible phase-transitions. It turns out that once the stability lost the black-hole never come back to its original stable position. 
\end{abstract}

\newpage
\setcounter{page}{1}
\pagestyle{plain}

\section{Introduction}
\label{sec:intro}

The solutions of Black-hole (BH) are natural outcome of the General theory of Relativity(GTR). Generally the BH characterized by their radius, masses, charges, angular momentum etc. Depending upon the parameter choices BH could have very different types, e.g., the  Schwarzschild, Kerr, Kerr-Newman, Reissner Nordstr\"{o}m\cite{Jacobson:2012ei}, \cite{Curiel:2018cbt}
. In our analysis we considered the Kerr solutions, i.e., the BH with rotation parameters.

The evolution of the different composite stars estimates that the rotation parameters could have varied ranges of 0.01 to 0.60 with very different characteristics features \cite{Miller:2014aaa}. The nearly static in our consideration is with spin parameter $a$ is 0.1 or less.\\

Nonlinear electrodynamics (NLED) has been attracted attention to astrophysicists since sometime back. NLED could have combined with General theory of relativity(GTR) -- we called as NLED-GTR. This theory has some remarkable features, e.g., consistent theory of inflationary model of the Universe \cite{Garcia,Camara,Elizalde,Novello,Novello1,Vollick,Kruglov3}.

 This NLED-GTR model has few characteristic features, e.g., absence of initial singularities, putting  an upper limit on the Electric field at the origin of point-like particles, the finite self-energy of the charged particles \cite{Novello1}, \cite{Born} \cite{Kruglov1}. 
 
 It is to be noted that in Quantum Electrodynamics(QED) the non-linear terms arises due to the loop corrections \cite{Heisenberg, Schwinger, Adler}. The NLED-GTR model follows the correspondence principle, i.e., in the weak-field limit we get the linear electrodynamics. 

The physics of BH (both electrically and magnetically charged) in presence of NLED has been studied since some time \cite{Kruglov1}, \cite{Bardeen}, \cite{Dymnikova}. We have studied the thermodynamics properties of the slowly rotating magnetized Kerr BH in NLED-GTR scenario.

 The paper is organized as follows: we briefly outline the NLED-GTR model in section 2. In section 3 we discussed the Kerr BH relevant for the NLED-GTR and estimated the magnetized mass density that affects the relevant thermodynamic parameters for slowly rotating (with spin parameter $a$) for $\gamma=\frac{1}{2}, \frac{4}{5}$ and $1$. We conclude in Section 6. 

We adopted natural units, i.e., $c=\hbar=1$, $\varepsilon_0=\mu_0=1$ and with metric signature $\eta=\mbox{diag}(-1,1,1,1)$


\section{ Non-linear Electrodynamics (NLED)-model}

The Lagrangian density is as (following \cite{Kruglov1}):
\begin{equation}
{\cal L} = -\frac{{\cal F}}{1+(\beta{\cal F})^\gamma},
 \label{lag1}
\end{equation}
where the parameter $\beta$ has the dimensions of 4D-volume (manifests the strength of the coupling), and $\gamma$ (manifests the order of interaction) is the dimensionless parameter \cite{Kruglov1}. The ${\cal F}=(1/4)F_{\mu\nu}F^{\mu\nu}=(\textbf{B}^2-\textbf{E}^2)/2$, where 
$F_{\mu\nu}=\partial_\mu A_\nu-\partial_\nu A_\mu$ is the field strength tensor, with $A_{\mu}= (\phi, \vec A)$ the $\phi$ and $\vec A$ are the scalar and vector potential respectively.

One can obtain the field equations \begin{equation}
\partial_\mu\left({\cal L}_{\cal F}F^{\mu\nu} \right)=0,
\label{lag2}
\end{equation} 
where ${\cal L}_{\cal F}=\partial {\cal L}/\partial{\cal F}$ by using the Euler-Lagrange (EL) equations.

The EL equation on Lagrangian density leads to \begin{equation}\label{lag3}
  {\cal L}_{\cal F}=\frac{(\gamma-1)(\beta{\cal F})^\gamma-1}{(1+(\beta{\cal F})^\gamma)^2}.
\end{equation}

The group velocity over the background is less than photon speed requires $ {\cal L}_{\cal F}\leq 0$ \cite{Shabad2} so that the causality principle holds and confirm the absence of tachyon. In our analysis we imply that $0\leq \gamma \leq 1$ and we consider only magnetized black holes (\textbf{E}=0, ${\cal F}=\textbf{B}^2/2$).
For simplicity in our analysis we set $\beta=1$.

The symmetrical energy-momentum tensor (is obtained from Eqn. \ref{lag1} ) is: 

\begin{equation}
T_{\mu\nu}=\frac{(\gamma-1)(\beta{\cal F})^\gamma-1}{[1+(\beta{\cal F})^\gamma]^2} F_\mu^{~\alpha}F_{\nu\alpha}
-g_{\mu\nu}{\cal L}.
\label{trace1}
\end{equation}

The trace of the energy-momentum tensor Eqn.\ref{trace1} leads to 

\begin{equation}\label{10}
  {\cal T}\equiv T^{\mu}_\mu=\frac{4\gamma {\cal F}(\beta{\cal F})^\gamma}{[1+(\beta{\cal F})^\gamma]^2}.
\end{equation}

The trace of Eqn.\ref{trace2} is non-zero due to the presence of dimensional parameter $\beta=1$(as we considered throughout our analysis) and it leads the breaking 
of scale invariance.

The models exhibits Maxwell's Electrodynamics at the weak field limit, $\beta {\cal F} \ll 1$, ${\cal L}\rightarrow-{\cal F}$, i.e., holds the correspondence principle.

\section{Magnetized Kerr black holes}

We consider the electromagnetic energy density to be small enough that the Kerr vacuum solution is valid. We have also considered the rotation parameter is small enough (consistent with astrophysical evidence of evolution of stars) 
that the Kerr-solution is almost nearly static (with rotation parameter $a \le 0.1$) over the magnetized NLED background.

The geometry of Kerr space-time metric is  
\begin{eqnarray}
 ds^2 & = & -dt^2 + \frac{2 M r}{\Sigma} \left( dt -a \sin^2 \theta d \phi \right)^2 
 + \frac{\Sigma}{\Delta} dr^2 + \Sigma d \theta^2  \nonumber \\
 & &+ (r^2 +a^2) \sin \theta d \phi^2\ ,
\end{eqnarray}
where
\begin{equation}
 \Sigma = r^2 + a^2 \cos^2 \theta , ~~~ \Delta = r^2 + a^2 - 2Mr .
\end{equation}

Here $M$ is the mass-energy parameter of the BH and $a$ is its rotation parameter. The thermodynamic quantities of a Kerr BH can be expressed in terms of its 
horizon radius $r_{+} = M + \sqrt{M^2 - a^2}$, which is defined by taking $\Delta = 0$.

The magnetic mass of the Kerr BH is 
\begin{equation}
M(r)=\int_0^r\rho_M(r)r^2dr,
\label{magmass}
\end{equation}
where $\rho_M$ is the magnetic energy density \cite{Kruglov1}. The magnetic mass of the BH is $m_M=\int_0^\infty\rho_M(r)r^2dr$.
The magnetic energy density (\textbf{E}=0), found from Eq.\ref{trace2}, is
\begin{equation}\label{17}
  \rho_M=T_0^{~0}=\frac{{\cal F}}{1+(\beta{\cal F})^\gamma}.
\end{equation}

For the magnetized black hole \cite{Bronnikov} one can find the field-strength with magnetic charge($q$) and at distance r is ${\cal F}=q^2/(2r^4)$. We set for simplicity the value of $q=1$.

Similar to \cite{Kruglov1} we have also considered $\gamma=\frac{1}{2}$. We have cross-checked our results with rotation parameter $a=0$ and found consistent with results in \cite{Kruglov1}.  

\section{Thermodynamic Parameters} In this section we are summarizing few important thermodynamic parameters \cite{Czinner:2017tjq} relevant for our analysis.\\

$\bullet$ The Bekenstein-Hawking entropy is 
\begin{equation}
 S_{BH} = \pi (r_{+}^2 + a^2),
\end{equation}

$\bullet$ The Hawking temperature of the black hole horizon is
\begin{equation}
 T_H = \frac{1}{2\pi}\left[ \frac{r_{+}}{r_{+}^2 + a^2} - \frac{1}{2 r_{+}} \right],
\end{equation}

$\bullet$ The angular momentum($J$) of the black hole is
\begin{equation}
 J = \frac{a}{2 r_{+}}(r_{+}^2 + a^2),
\end{equation}

$\bullet$ The specific heat capacity at constant angular momentum is
\begin{equation}
 C_J = T_H \left( \frac{\partial S_{BH}}{\partial T_H} \right)_J 
 = \frac{2 \pi (r_{+}^2 -a^2)(r_{+}^2 + a^2)^2}{3 a^4 + 6 r_{+}^2 a^2 - r_{+}^4}\ .
\end{equation}

\section{Numerical analysis}

In our numerical analysis we have considered three different values of $\gamma=\frac{1}{2}, \frac{4}{5}$ and $1$. 

We assume the value of spin parameter ($a$) is slowly rotating or nearly static. Let us comment briefly on that. Some massive stars especially stars with initial masses $M> 20M_{Sun}$, produce BHs upon core-collapse. The resulting stellar remnant is a rotating Kerr BH, whose
dimensionless spin is defined as  $a=\frac{Jc}{GM^2}$, where J is the angular momentum of the BH, c is the speed of light, G is Newton's Gravitational constant and M is the mass of the BH ~\cite{Fuller:2019sxi}. The values of spin parameter ($a$) can be in wide ranges: from 0.006 to 0.549 ~\cite{Fuller:2019sxi}. The numerical values of $a=0.1 (0.01)$ could be possible for a BH formed from a single massive star irrespective of the angular momentum.  The BH speed rates are measured in X-ray binaries (XRBs), and current estimates suggest a broad range of spin-rates $0.1 \le a \le 1$  \cite{Miller:2014aaa}. 

\subsection{$\gamma=\frac{1}{2}$}

\begin{figure}[ht!]
\begin{center}
\raisebox{0.0cm}{\hbox{\includegraphics[angle=0,scale=0.33]{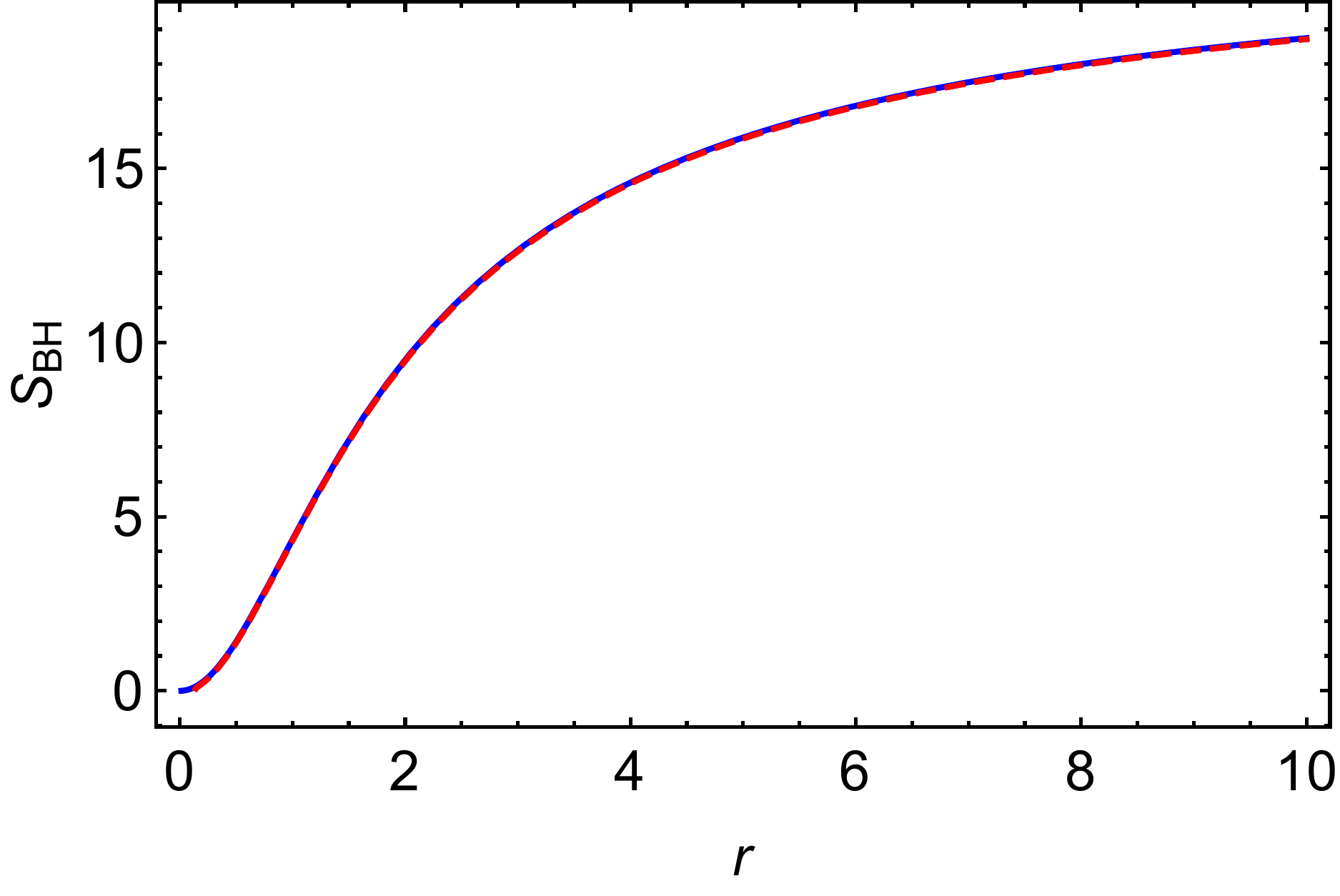}}}
\raisebox{0.0cm}{\hbox{\includegraphics[angle=0,scale=0.33]{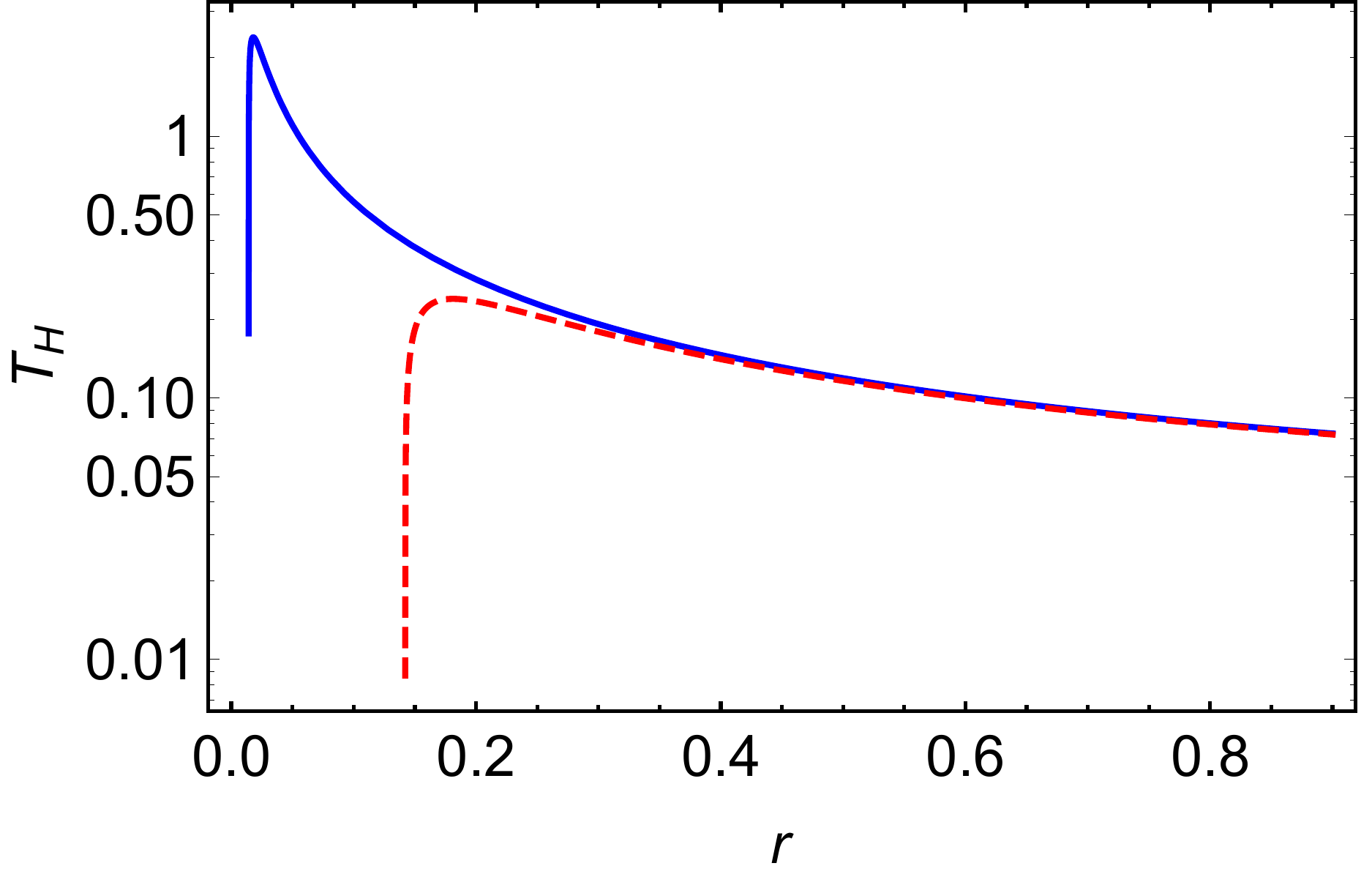}}}
\caption{ The Entropy (left) and Hawking temperature (right) for $\gamma=\frac{1}{2}$
}
\label{fig12a}
\end{center}
\end{figure}

We have plotted the entropy (in the left-panel of  Fig.\ref{fig12a} ) as a function of radial co-ordinates. The entropy is almost identical for both of the two-spin parameters under consideration.

We have plotted the Hawking temperature (right-panel of Fig.\ref{fig12a}) as a function of radial co-ordinates. It seems that the Hawking temperature is positive approximate at $r \ge$ 0.02 (0.2) for spin parameter $a=0.01(0.1)$ with solid-blue(dashed-red).

\begin{figure}[ht!]
\begin{center}
\raisebox{0.0cm}{\hbox{\includegraphics[angle=0,scale=0.33]{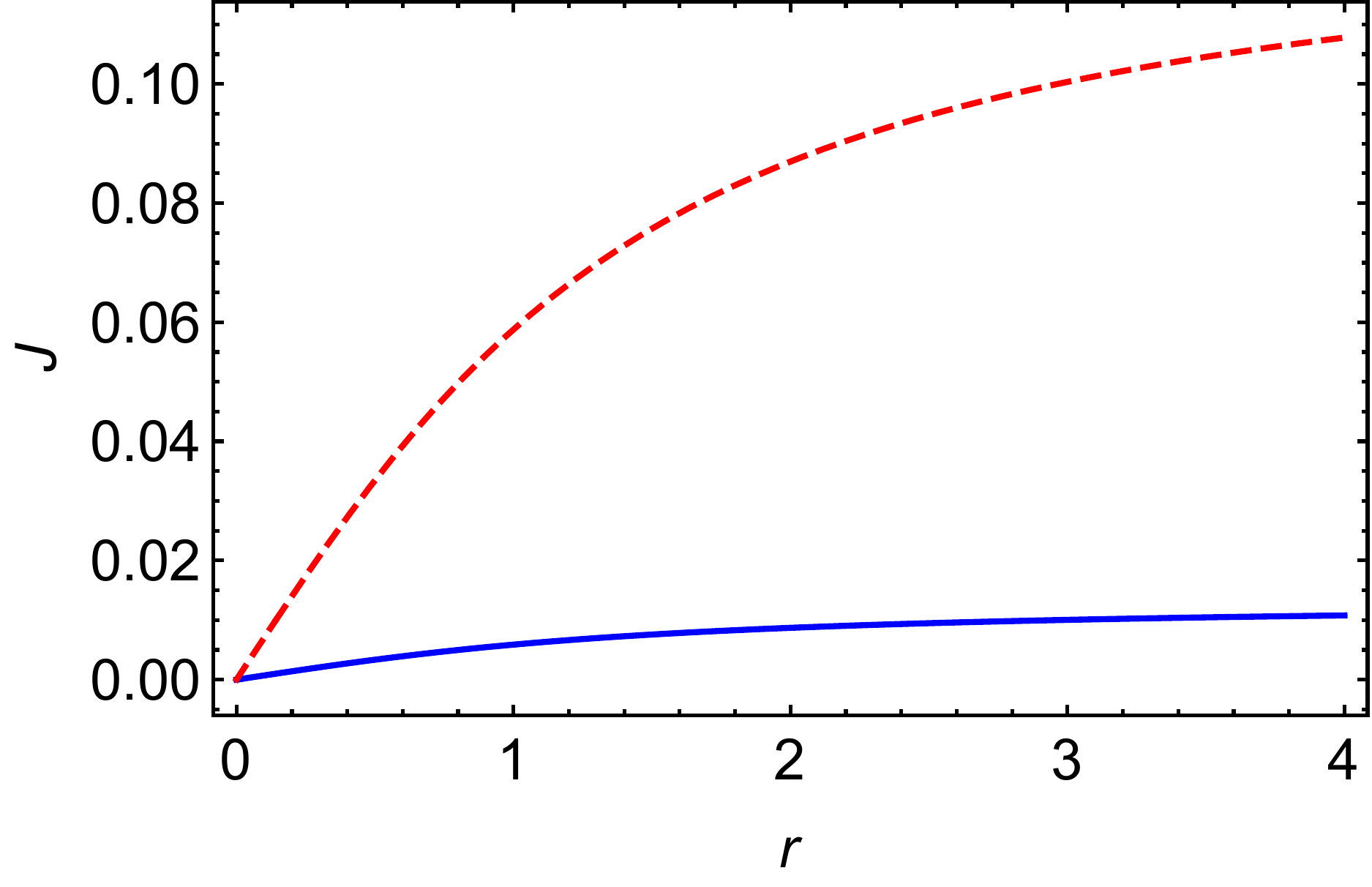}}}
\raisebox{0.0cm}{\hbox{\includegraphics[angle=0,scale=0.20]{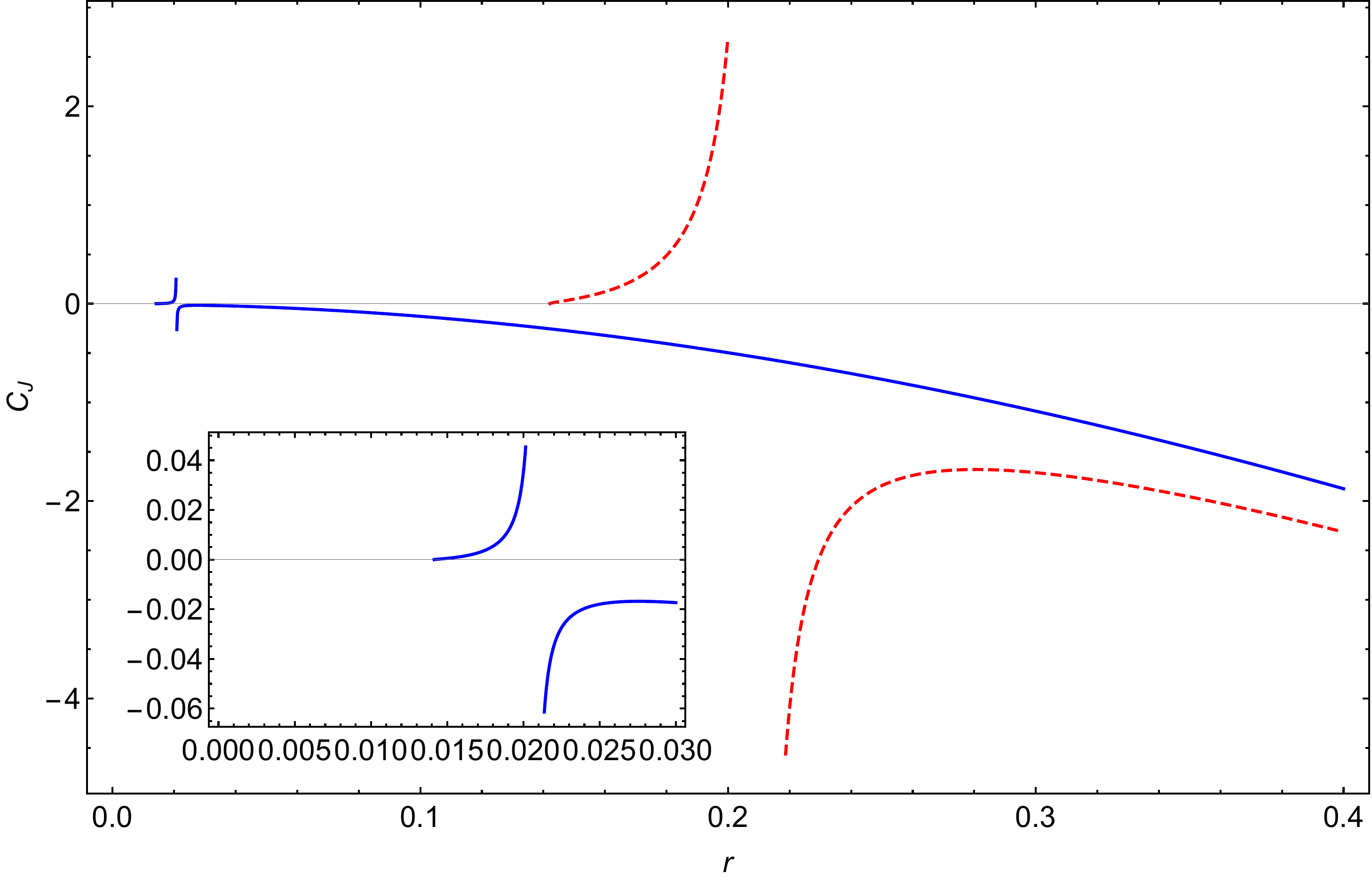}}}
\caption{ The angular momentum($J$) and the specific heat at constant angular momentum  ($C_J$) for slowly rotating Kerr BH (with spin parameter, $a$=$0.01$ with  solid-blue and $0.1$ with dashed-red) in non-linear electrodynamics. The value of $\gamma$ is $\frac{1}{2}$}
\label{fig12b}
\end{center}
\end{figure}

In the left(right)-panel of Fig.\ref{fig12b} we have plotted the angular momentum (specific heat at constant angular momentum). It reflects the fact that with low(high) spin parameter the angular momentum is small(large). The specific heat 
is discontinuous at $r=0.02(0.2)$ for the spin parameter $a=0.01(0.1)$ and the BH undergoes second order phase-transition. In this region of r the Heat capacity is negative and the BH is unstable. Moreover for larger r the specific heat ($C_J$) is always negative it suggests that once the BH undergoes instability it never comes back to its original stable position.

\subsection{$\gamma=\frac{4}{5}$}

\begin{figure}[ht!]
\begin{center}
\raisebox{0.0cm}{\hbox{\includegraphics[angle=0,scale=0.33]{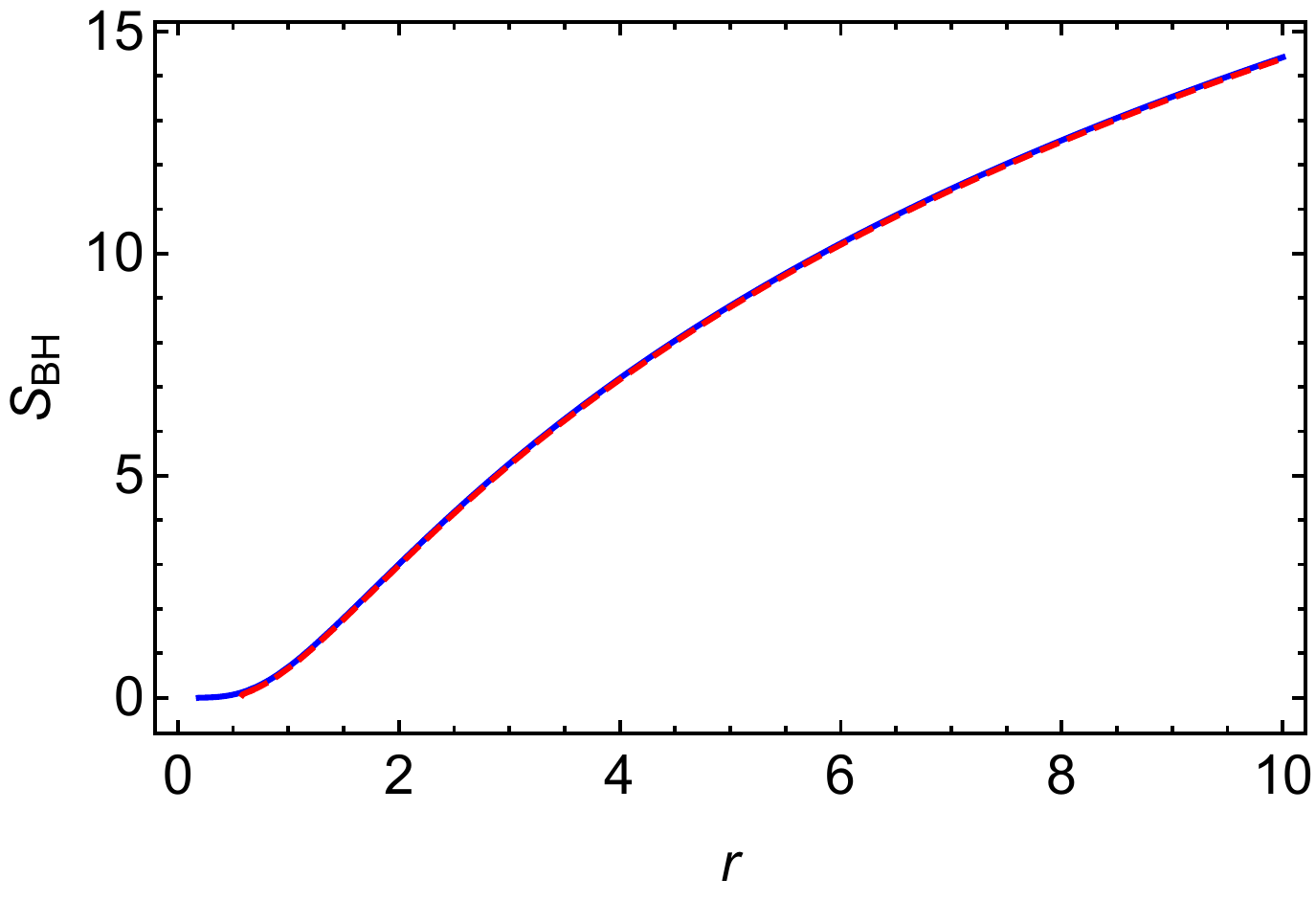}}}
\raisebox{0.0cm}{\hbox{\includegraphics[angle=0,scale=0.33]{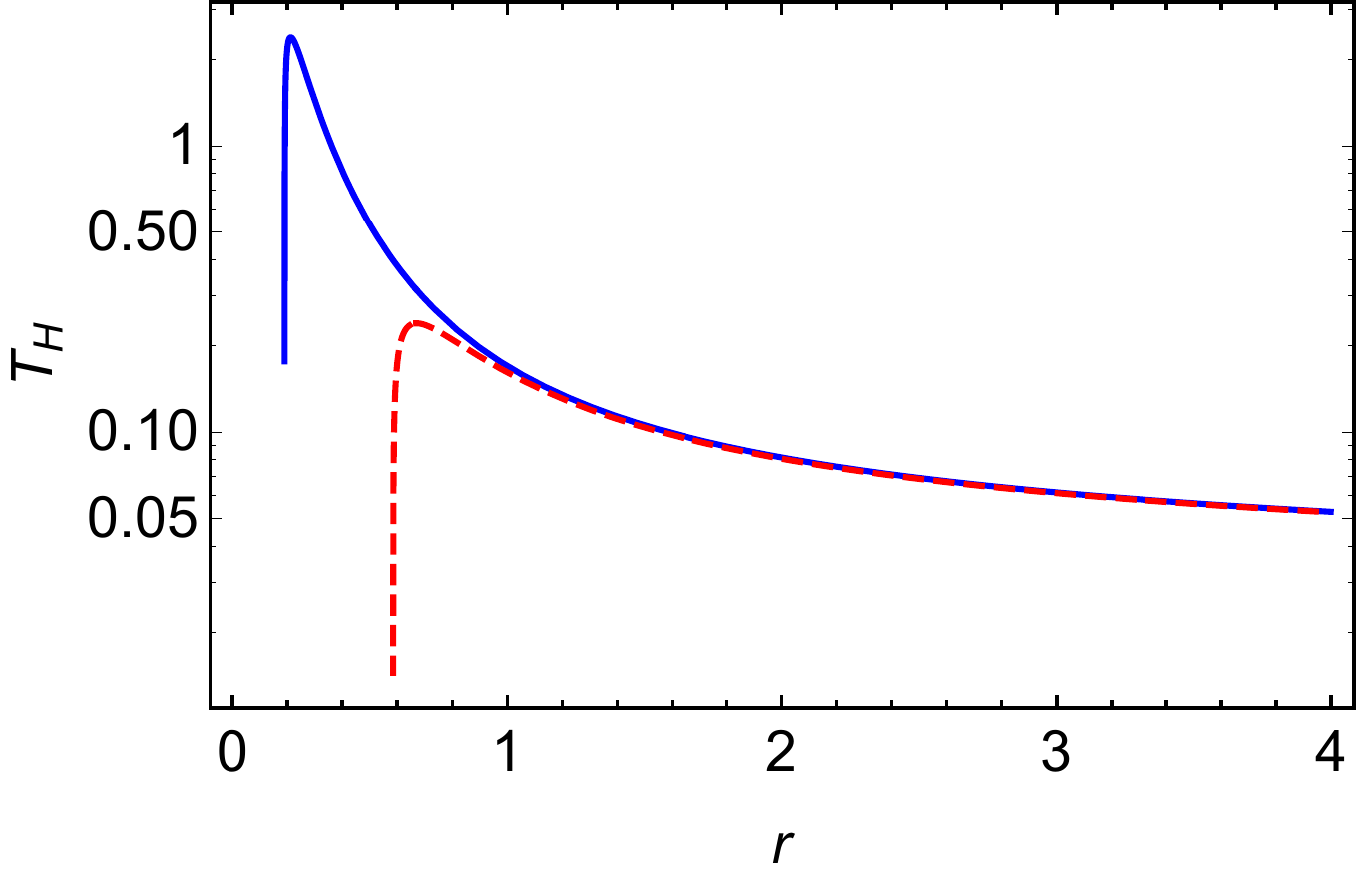}}}
\caption{ The Entropy (left) and Hawking temperature (right) for $\gamma=\frac{4}{5}$
}
\label{fig45a}
\end{center}
\end{figure}

We have plotted the Entropy (in the left-panel of  Fig.\ref{fig45a} ) as a function of radial co-ordinates for $\gamma=\frac{4}{5}$. The entropy is almost identical for both of the two-spin parameters. The values of spin parameter for nearly static Kerr solutions are small enough to show-up any appreciable changes.

We have plotted the Hawking temperature (right-panel of Fig.\ref{fig45a}) as a function of radial co-ordinates. It seems that the Hawking temperature is positive approximate at $r \ge$ 0.225 (0.7) for spin parameter $a=0.01(0.1)$ with solid-blue (dashed-red). It turns out that the large values of $\gamma$ with respect to Fig.\ref{fig12a} the BH instability occurs for larger values of radial parameters. 

\begin{figure}[ht!]
\begin{center}
\raisebox{0.0cm}{\hbox{\includegraphics[angle=0,scale=0.33]{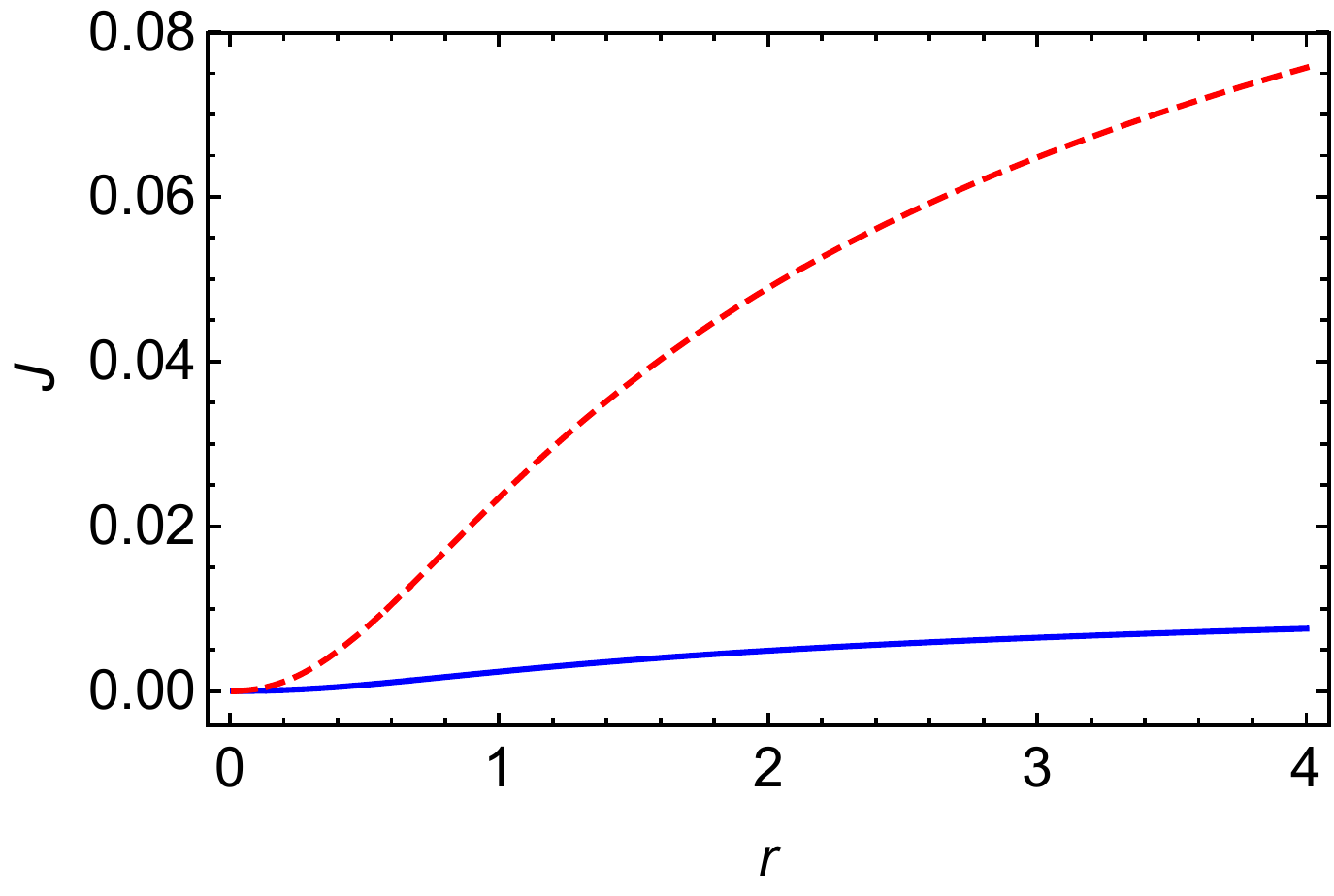}}}
\raisebox{0.0cm}{\hbox{\includegraphics[angle=0,scale=0.20]{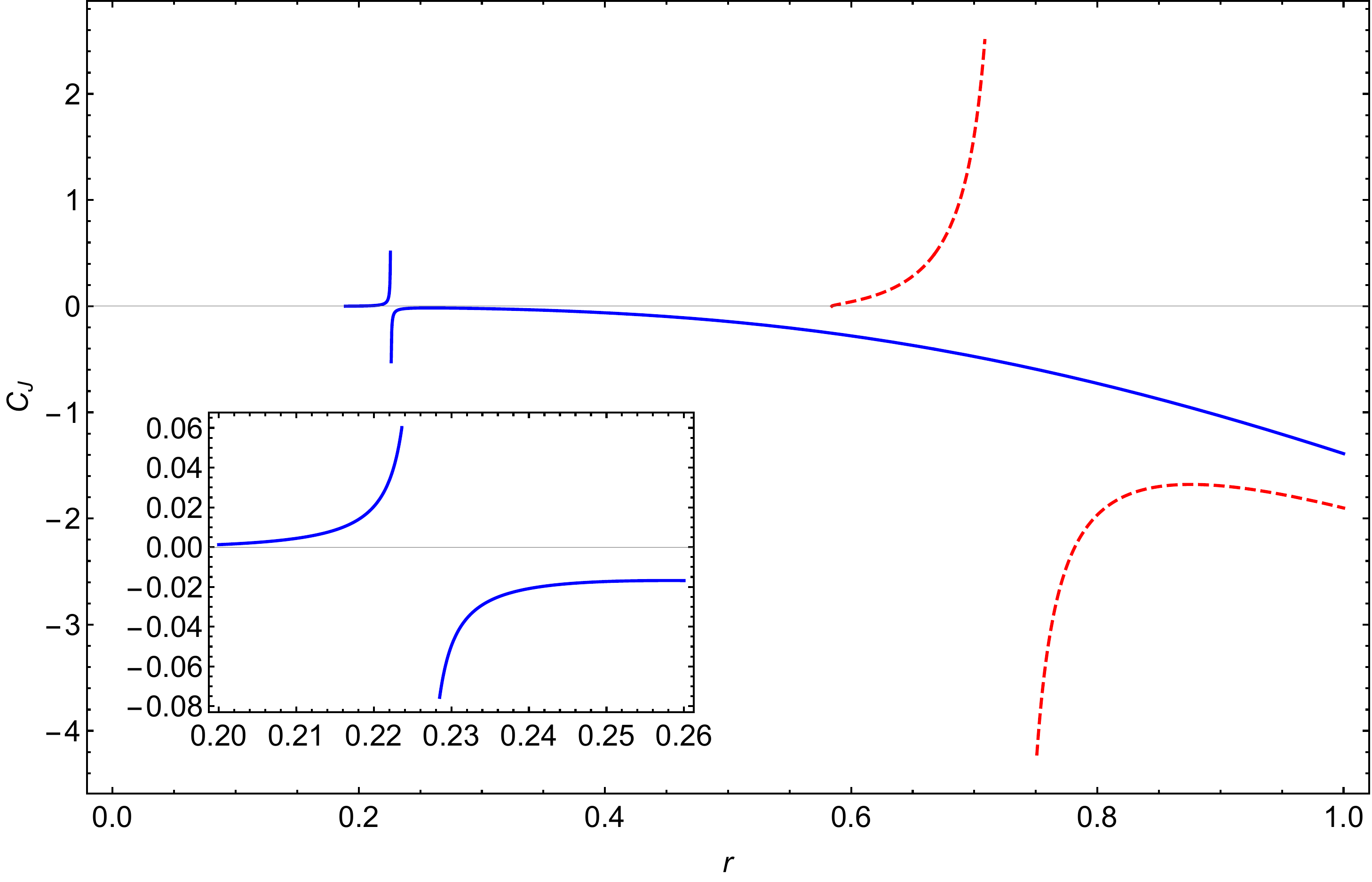}}}
\caption{ The angular momentum($J$) and the specific heat at constant angular momentum  ($C_J$) for slowly rotating Kerr BH (with spin parameter, $a$=$0.01$ with  solid-blue and $0.1$ with dashed-red) in non-linear electrodynamics. The value of $\gamma$ is $\frac{4}{5}$}
\label{fig45b}
\end{center}
\end{figure}

In the left(right)-panel of Fig.\ref{fig45b} we have plotted the angular momentum (specific heat at constant angular momentum). It reflects the fact that with low(high) spin parameter the angular momentum is small(large). The specific heat 
is discontinuous at $r=0.0225(0.7)$ for the spin parameter $a=0.01(0.1)$ and the BH undergoes second order phase-transition. In this region of radial distance the specific heat capacity ($C_J$) is negative and the BH is unstable.

\subsection{$\gamma=1$}

\begin{figure}[ht!]
\begin{center}
\raisebox{0.0cm}{\hbox{\includegraphics[angle=0,scale=0.35]{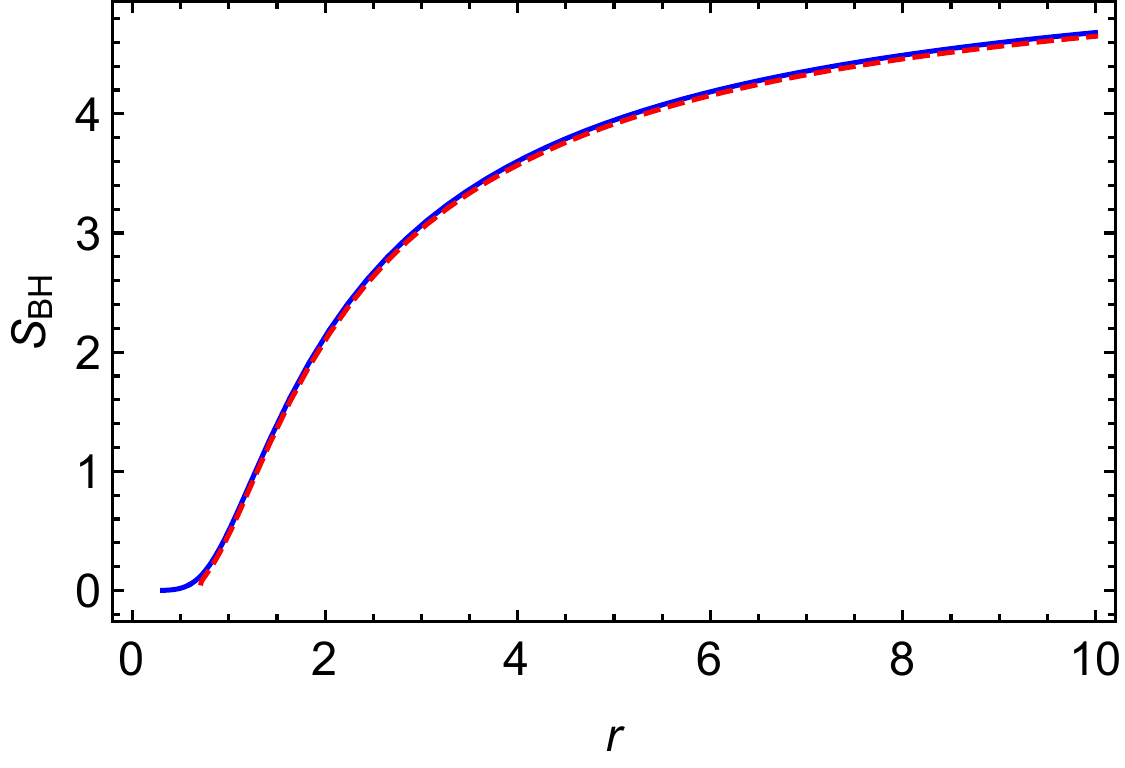}}}
\raisebox{0.0cm}{\hbox{\includegraphics[angle=0,scale=0.35]{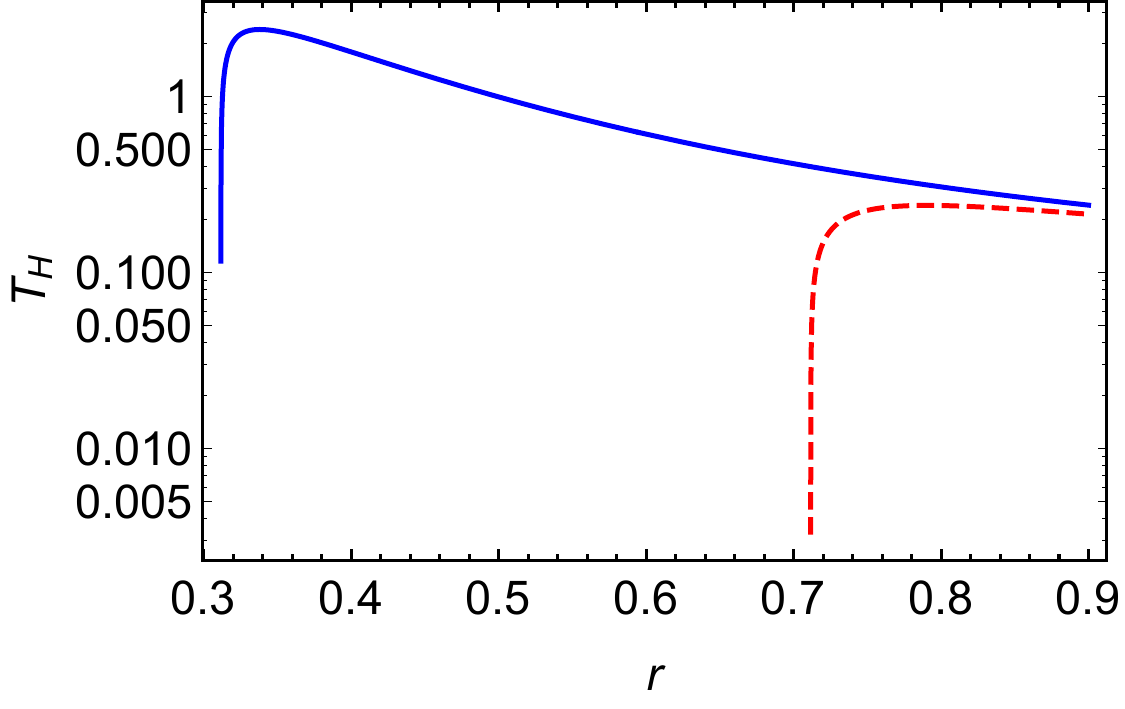}}}
\caption{ The Entropy (left) and Hawking temperature (right) for $\gamma=1$
}
\label{fig1a}
\end{center}
\end{figure}

\begin{figure}[ht!]
\begin{center}
\raisebox{0.0cm}{\hbox{\includegraphics[angle=0,scale=0.35]{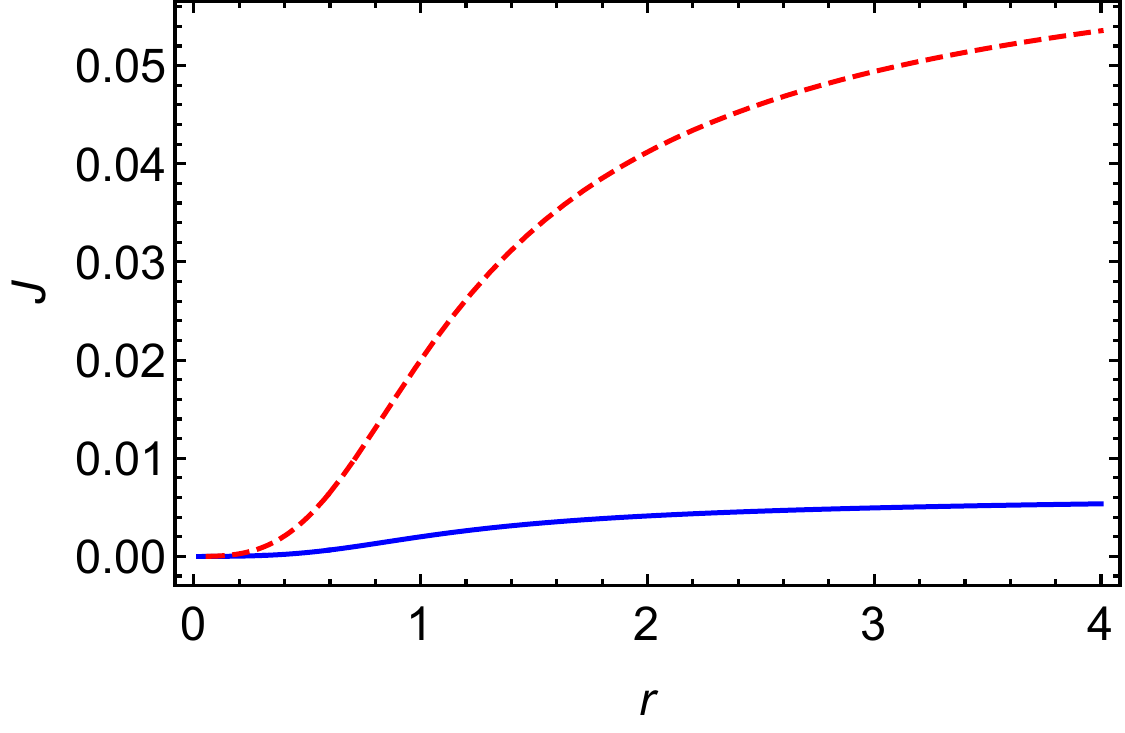}}}
\raisebox{0.0cm}{\hbox{\includegraphics[angle=0,scale=0.21]{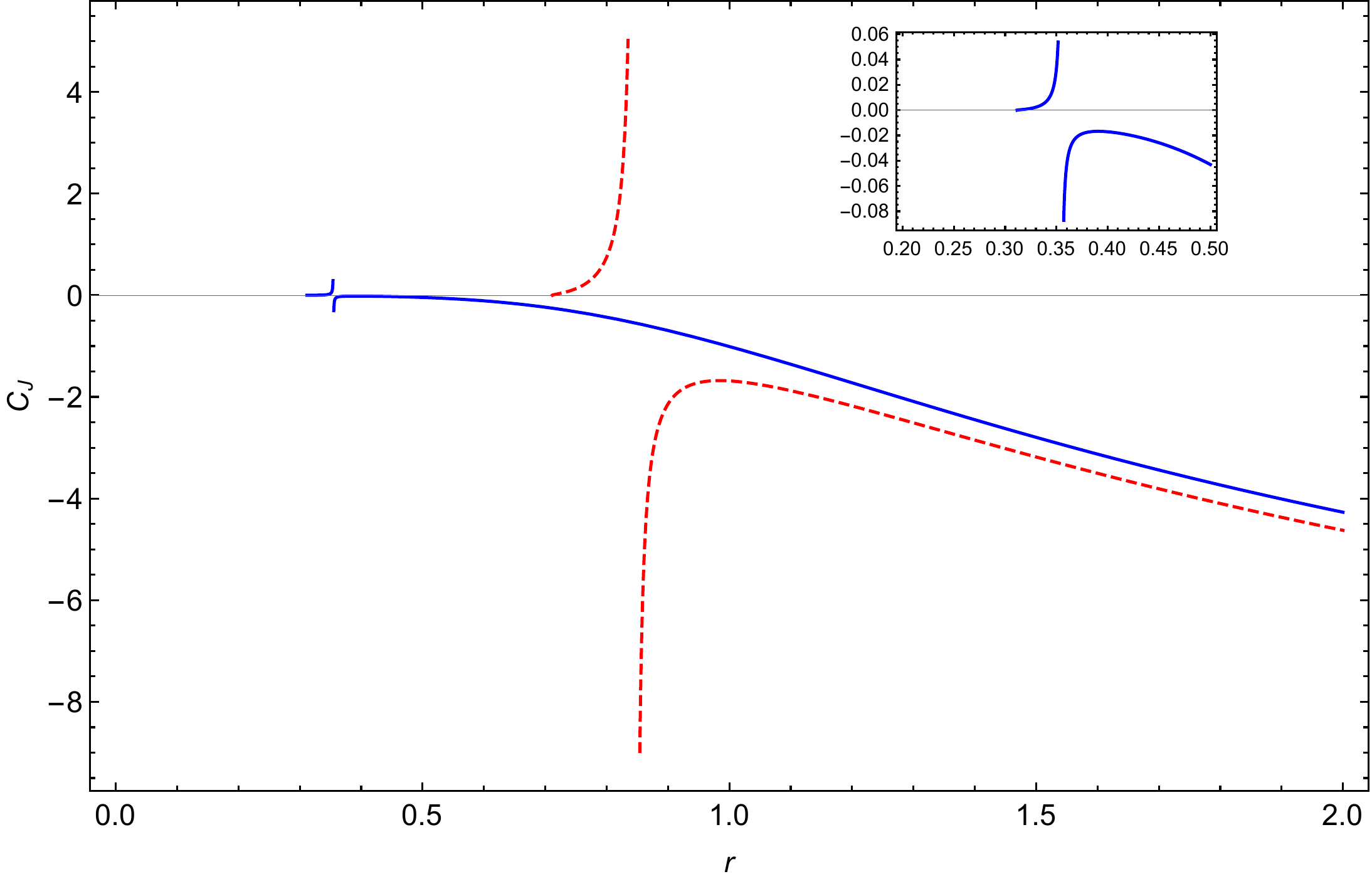}}}
\caption{ The angular momentum($J$) and the specific heat at constant angular momentum  ($C_J$) for slowly rotating Kerr BH (with spin parameter, $a$=$0.01$ with  solid-blue and $0.1$ with dashed-red) in non-linear electrodynamics. The value of $\gamma$ is $1$.}
\label{fig1b}
\end{center}
\end{figure}

The allowed values of $\gamma$ is $ 0 \le \gamma \le 1 $ and we considered the extreme value. We have plotted the Entropy (in the left-panel of  Fig.\ref{fig1a} ) as a function of radial co-ordinates. The entropy is almost identical for both of the two-spin parameters. As the values of spin parameter for near static Kerr solutions are small enough to show-up any appreciable changes.

We have plotted the Hawking temperature (right-panel of Fig.\ref{fig1a}) as a function of radial co-ordinates. It seems that the Hawking temperature is positive approximate at $r \ge$ 0.31 (0.71) for spin parameter $a=0.01(0.1)$ with solid-blue (dashed-red). It seems that the large values of $\gamma$ with respect to Fig.\ref{fig12a} the BH instability occurs for the large values of radial parameters.

In the left(right)-panel of Fig.\ref{fig1b} we have plotted the angular momentum (specific heat at constant angular momentum). It reflects the fact that with low(high) spin parameter the angular momentum is small(large). The specific heat 
is discontinuous approximately at $r=0.31(0.71)$ for the spin parameter $a=0.01(0.1)$ and the BH undergoes second order phase-transition. For larger r the specific heat capacity ($C_J$) never comeback to positive values so we can say that BH is unstable.

\section{Summary and Conclusion}
We have analyzed the thermodynamic properties of magnetized Black-hole in presence of the non-linear electrodynamics with two independent parameters 
$\beta$ and $\gamma$. In our numerical analysis we set $\beta=1$ throughout. We assumed the magnetized Black-hole is rotating very slowly 
with two values of rotation parameter ($a$) 0.01 and 0.1 (both normalized to mass), so that nearly static Kerr metric could be used for the magnetized BH. The correspondence principle takes place so that for the non-rotational model the metric is converted into the Schwarzschild metric in the vanishing limits of $a$-parameter. We have analyzed several thermodynamic  parameters, for example Black-hole entropy, Hawking temperature, angular momentum and specific heat capacity at constant angular momentum. We have also demarcated the regions where the Black-holes undergoes second-order phase transitions and regions of instability.


\end{document}